\documentclass[AMA,STIX1COL]{WileyNJD-v2}

\articletype{Research Article}%

\received{DD Month YYYY}
\revised{DD Month YYYY}
\accepted{DD Month YYYY}

\usepackage{enumitem}
\usepackage{bbm}

\newcommand{\bs}{\mbox{\boldmath $s$}}

\def\bs{\boldsymbol}

\newcommand{\indep}{\rotatebox[origin=c]{90}{$\models$}}

\newcommand{\E}{\mathbb{E}}

\newcommand{\ind}[1]{\mathbbm{1}_{#1}}

\begin{document}

%\title{This is the sample article title\protect\thanks{This is an example for title footnote.}}

\title{A Framework for Mediation Analysis with Multiple Exposures, Multivariate Mediators, and Non-Linear Response Models}

\author[1]{James P. Long}

\author[1]{Ehsan Irajizad}

\author[2]{James D. Doecke}

\author[1]{Kim-Anh Do}

\author[1]{Min Jin Ha*}

\address[1]{\orgdiv{Department of Biostatistics}, \orgname{University of Texas MD Anderson Cancer Center}, \orgaddress{\state{Texas}, \country{USA}}}

\address[2]{\orgdiv{CSIRO}, \orgname{Royal Brisbane and Women’s Hospital}, \orgaddress{\state{Brisbane}, \country{Australia}}}

%\address[3]{\orgdiv{Org Division}, \orgname{Org Name}, \orgaddress{\state{State name}, \country{Country name}}}

\corres{*\email{mjha@mdanderson.org}}

%\presentaddress{This is sample for present address text this is sample for present address text}

\abstract[Summary]{Mediation analysis seeks to identify and quantify the paths by which an exposure affects an outcome. Intermediate variables which are effected by the exposure and which effect the outcome are known as mediators. There exists extensive work on mediation analysis in the context of models with a single mediator and continuous and binary outcomes. However these methods are often not suitable for multi--omic data that include highly interconnected variables measuring biological mechanisms and various types of outcome variables such as censored survival responses. In this article, we develop a general framework for causal mediation analysis with multiple exposures, multivariate mediators, and continuous, binary, and survival responses. We estimate mediation effects on several scales including the mean difference, odds ratio, and restricted mean scale as appropriate for various outcome models. Our estimation method avoids imposing constraints on model parameters such as the rare disease assumption while accommodating continuous exposures. We evaluate the framework and compare it to other methods in extensive simulation studies by assessing bias, type I error and power at a range of sample sizes, disease prevalences, and number of false mediators. Using Kidney Renal Clear Cell Carcinoma data from The Cancer Genome Atlas, we identify proteins which mediate the effect of metabolic gene expression on survival. Software for implementing this unified framework is made available in an R package (\url{https://github.com/longjp/mediateR}).}

%\keywords{keyword1, keyword2, keyword3, keyword4}

\keywords{Mediation Analysis, Direct Effects, Indirect Effects, Non--linear models, Multiomics}

% \jnlcitation{\cname{%
% \author{Williams K.}, 
% \author{B. Hoskins}, 
% \author{R. Lee}, 
% \author{G. Masato}, and 
% \author{T. Woollings}} (\cyear{2020}), 
% \ctitle{A regime analysis of Atlantic winter jet variability applied to evaluate HadGEM3-GC2}, \cjournal{Q.J.R. Meteorol. Soc.}, \cvol{2017;00:1--6}.}

\maketitle

%\footnotetext{\textbf{Abbreviations:} ANA, anti-nuclear antibodies; APC, antigen-presenting cells; IRF, interferon regulatory factor}

\section{Introduction} 
\label{sec:intro}
%Mediation analysis seeks to identify and quantify the paths by which one quantity causes changes in another. For example, if the expression level of a gene causes changes in the expression level of a protein which subsequently impacts survival, then the protein is a mediator of the gene -- survival relation.

In recent decades, numerous technological advancements have provided the ability to deeply characterize the molecular properties of tissues at different levels, i.e., genomics, transcriptomics, proteomics and epigenetics. These levels form a hierarchical structure in which alterations at one level have the capacity to cause changes downstream. For example, the central dogma of biology states that information flows from mRNA to proteins via translation. Thus the effects of changes at the mRNA level on a phenotype such as survival may be \textit{mediated} by changes in protein expression \citep{kumar2016integrating}. Identification and quantification of such mediators enhances scientific understanding of how changes at one level impact a phenotype. Mediators offer targets for therapeutic intervention in the case of a disease phenotype.

The literature on mediation analysis dates back to \citet{baron1986moderator}, who studied the concept in linear models with a single mediator. \cite{robins1992identifiability} and \cite{pearl2001direct} generalized the definitions of direct and indirect effects to include non--linear models. Since then, estimation of mediation effects has been studied with various outcome distributions \citep{imai2010general}, with multiple mediators \citep{huang2016hypothesis,fasanelli2019marginal,zhao2020sparse}, and on different effect scales \citep{vanderweele2010odds}. 

Existing modeling frameworks have limitations. \cite{huang2014joint} and \cite{vanderweele2010odds} proposed mediation methods with logistic response models but require the response to be rare. \cite{gaynor2018mediation} proposed a probit approximation to the logistic function, suitable only for common responses. For survival responses, propensity models have been proposed which can only accommodate binary or categorical exposures. However these will not work with continuous exposures such as mRNA expression \citep{fasanelli2019marginal}. Other frameworks such as \cite{imai2010general} accommodate a wide range of response models but measure effects only on the mean difference scale, which is often not appropriate for binary or survival responses.

This work makes several methodological advances which extend the statistical models and causal structures to which mediation analysis can be applied. These developments are particularly relevant to multiomics data sets which contain multiple potential causes (e.g. many gene expression measures), multiple mediators (e.g. many proteins), and responses not suitable for linear models (such as survival time or disease status). Our framework 1) estimates mediation effects with vector valued mediators without requiring specification of the causal structure among the mediators 2) handles Gaussian, logistic, and survival response models while measuring mediation effect on various scales appropriate to the given response model and 3) eliminates restrictive assumptions such as requiring binary exposures or ``rare diseases''. A publicly available R package incorporating all of this functionality facilitates use of this framework by others.

In this work we apply our mediation framework to the Kidney Renal Clear Cell Carcinoma (KIRC) project of The Cancer Genome Atlas (TCGA), a multiomics data set consisting of genomic, transcriptomic, proteomic, and clinical data. \cite{cancer2013comprehensive} identified metabolic genes and proteins which correlate with survival in KIRC. We use mediation analysis to investigate how shifts in metabolic pathways at the gene expression level change survival by altering the expression levels of multiple metabolic proteins, and other key proteomic pathways.

%TCGA generated multi--platform molecular profiles (including gene and protein expression) and large clinical datasets including curated survival endpoints of more than 11,000 human tumors across 33 different cancer types and subtypes \url{(http://cancergenome.nih.gov)} \citep{liu2018integrated}. We consider kidney renal clear cell carcinoma (KIRC), the most common and lethal type of kidney cancer.We are interested in evaluating the effect of pathway activities calculated at the gene expression level on patients' overall survival times, mediated by translational mechanisms of genes.
% , involving abnormal expression of genes and proteins related to PTEN, the citrate (TCA) cycle, Fatty acid synthesis (FAS), AMP-activated kinase (AMPK) and Pentose phosphate and acetyl-CoA carboxylase (ACC). 

This work is organized as follows: Section \ref{sec:frame} describes our framework in--depth and compares it with existing methods (Section \ref{sec:prior_work}). The value of our framework is demonstrated in simulations in Section \ref{sec:sim} and an application to the TCGA KIRC data set in Section \ref{sec:application}. We conclude with a discussion in Section \ref{sec:discussion}.

\section{Causal Mediation Analysis Framework}
\label{sec:frame}

\subsection{Causal Structure}

\begin{figure}[H]
  \centering
   \includegraphics[width=0.9\textwidth]{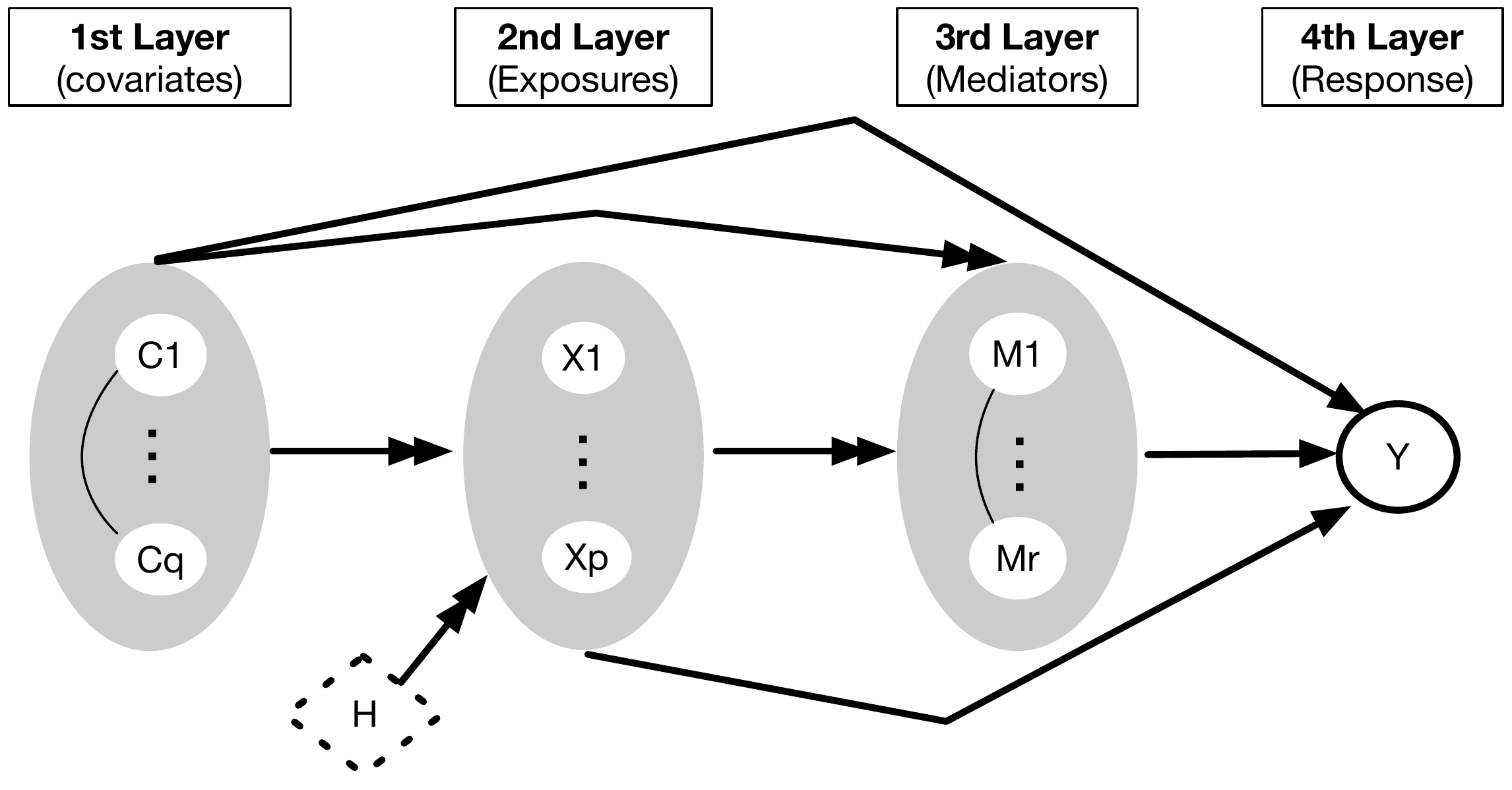}
\caption{Illustration of directed acyclic graph (DAG) for mediation analysis, where four disjoint sets of variables (nodes), covariates ($\bs{C}$), exposures ($\bs{X}$), mediators ($\bs{M}$) and $Y$ (response), have their unique order, $\bs{C}<\bs{X}<\bs{M}<Y$. The goal is to assess the causal impact of changing any single exposure $X\in \{X_1,\ldots X_p\}$  on an outcome $Y$ and quantify how much of this effect is mediated by the set of mediators $\bs{M}=\{M_1,\ldots,M_r\}$. The variables $\bs{C}=\{C_1,\ldots,C_q\}$ represent potential confounders. Our model assumes that the causal agents may be linked by unobserved factors ($H$) and permits mediators to have internal causal or correlation structure.\label{fig:dag}}
\end{figure}

Our mediation analysis framework is based on the causal directed acyclic graph (DAG) structure illustrated in Figure~\ref{fig:dag}. There are four data layers: covariates $\bs{C}=(C_1,\ldots,C_q)^T\in \mathbb{R}^q$, exposures $\bs{X} = (X_1,\ldots, X_p)^T\in \mathbb{R}^p$, mediators $\bs{M} = (M_1,\ldots,M_r)^T\in \mathbb{R}^r$, and outcome layer $Y$. Variables in each layer (e.g., $\bs{X}$) potentially have causal influence on variables in the downstream layers (e.g., $\bs{M}$ and $Y$ but not $\bs{C}$).

Our framework assesses the causal effect of any of the $X_i$ for $i=1,\ldots, p$ on the outcome $Y$ and quantifies how much of this effect passes through the set of mediators $\bs{M}$, termed indirect effect, and how much of the effect is through other mechanisms, termed direct effect \citep{pearl2001direct,pearl2009causality}.

We assume that the correlations among the exposures are the result of observed confounders $\bs{C}$ which may also confound the mediator--exposure relation and unobserved confounders $H$ which only causally influence $\bs{X}$ (not $\bs{M}$ or $Y$). Thus the causal assumptions imply $X_i \indep X_j | \bs{C},\bs{H}$. Note that the mediation effects for $X_i$ in our model cannot be derived by treating the other $X$ variables (termed $\bs{X}_{-i}$) as confounders since this would assume that $\bs{X}_{-i}$ are causes of $X_i$. 

%For example, $\bs{X}$ includes $p$ gene expression profiles, $Y$ a prognostic phenotype (e.g., overall survival), $\bs{M}$ $r$ protein expression profiles, and $\bs{C}$ includes $q$ genetic or environmental factors (e.g., single nucleotide polymorphisms, SNPs or air pollution). 

In the real data example (Section \ref{sec:application}), we investigate the effects of metabolic gene expression  $\bs{X}$ on overall survival $Y$ and the extent to which this effect is mediated by metabolic proteins $\bs{M}$ (translational mechanism as indirect effect) or occurs through unmeasured gene regulatory paths of $\bs{X}$ to the outcome (direct effect). In this example, $H$ would represent biological mechanisms that govern crosstalk among pathways \citep{sam2016xtalkdb}.

For each $X_i$ we consider a single indirect effect for the set of mediators $\bs{M}$, rather than attempting to assess the indirect effect of individual $M_j$. There are several reasons for this approach: 1) We avoid having to specify any internal causal structure among the $\bs{M}$ variables. 2) Path effects (effect of $X_i$ on $Y$ through only $M_j$) cannot be identified when there is a $M_j$, $Y$ confounder which is itself influenced by $X$, which is likely to be the case when $\bs{M}$ represents a set of measures on the same genomic platform \citep{avin2005identifiability,vanderweele2014effect} 3) In the context of genomic data where $\bs{M}$ represents a set of variables in a pathway observed on the same platform (e.g. metabolic protein expression in Section \ref{sec:application}), the indirect effect is the causal effect of $X_i$ on $Y$ which is jointly mediated by the entire pathway, a desirable interpretation.

\subsection{Counterfactual Random Variables and Assumptions}

Counterfactual random variables are used to formally define causal interventions and the notions of direct and indirect effects. Let $Y^{X=x'}$ be the value of $Y$ obtained by setting $X=x'$, possibly counter to fact. For notational simplicity we will write $Y^{x'}$ when it is clear $X$ is being set to $x'$. Counterfactual notation can also express interventions on multiple variables. For example $Y^{x',\bs{m}'}$ is the value $Y$ would obtain by setting $X=x'$ and $\bs{M}=\bs{m}'$. Direct and indirect effects are represented as functions of nested counterfactual such as $Y^{x'',\bs{M}^{x'}}$, the value $Y$ would have obtained had $X$ been set to $x''$ and $M$ been set to the value it would have obtained had $x$ been set to $x'$.

The following counterfactual independence and consistency relations are needed to express direct and indirect effects in terms of the joint distribution. For $X_i$ for $i=1,\ldots,p$:

\begin{enumerate}[label=(\roman*)]
\item $\bs{M}^{X_i=x'} \indep  X_i | \bs{C}, \bs{X}_{-i} \, \, \, \forall \, \, \, x'$ \label{assump:1}: The value $\bs{M}$ obtains when $X_i$ is set to $x'$ is independent of $X_i$ given  $\bs{C}$ and all other exposures $\bs{X}_{-i}$.
\item $Y^{X_i=x'',\bs{m}'} \indep \bs{M}^{X_i=x'} | \bs{C}, \bs{X}_{-i} \, \, \, \forall \, \, \, x',x'',\bs{m}'$  \label{assump:4}: The value that $Y$ obtains when $X_i$ is set to $x''$ and $\bs{M}$ is set to $\bs{m}'$ is independent of the value of $\bs{M}$ when it is set to the value it would take had $X_i$ been $x'$, given $\bs{C}$ and other exposures $\bs{X}_{-i}$. 
\item $Y^{X_i=x'',\bs{m}'} \indep X_i | \bs{C}, \bs{X}_{-i} \, \, \, \forall \, \, \, x'',\bs{m}'$ \label{assump:5}: The value that $Y$ obtains when $X_i$ is set to $x''$ and $\bs{M}$ is set to $\bs{m}'$ is independent of $X_i$, given $\bs{C}$ and other exposures $\bs{X}_{-i}$. 
\item  $Y^{X_i=x'',\bs{m}'} \indep \bs{M} | \bs{C}, \bs{X} \, \, \, \forall \, \, \, x'',\bs{m}'$ \label{assump:6}: The value that $Y$ obtains when $X_i$ is set to $x''$ and $\bs{M}$ is set to $\bs{m}'$ is independent of $M$, given $\bs{C}$ and other exposures $\bs{X}$. 
\item $X_i=x' \implies \bs{M}^{X_i=x'} = \bs{M} \, \, \, \forall \, \, \, x'$ \label{assump:2} \, \text{(consistency)}: If $X_i=x'$ then the value that $\bs{M}$ takes when setting $X_i=x'$ is $\bs{M}$.
\item $\bs{M} = \bs{m}', X_i=x' \implies Y^{X_i=x',\bs{m}'} = Y$ \label{assump:7} \, \text{(consistency)}: If $X_i=x'$ and $\bs{M}=\bs{m}'$, then the value that $\bs{Y}$ takes when setting $X_i=x'$ is $\bs{M}=\bs{m}'$ is equal to $Y$.
\item $Y^{X_i=x'} = Y^{X_i=x',\bs{m}^{X_i=x'}}  \, \, \, \forall \, \, \, x'$ \label{assump:8} \, \text{(composition)}
\end{enumerate}

These relations can be viewed as assumptions regarding the elemental counterfactual random variables, following the potential outcomes framework of \cite{rubin1974estimating}, or as consequences of the assumptions encoded in the causal DAG structure, following the approach of \cite{pearl2009causal} (Section 7.3 p. 228). We refer to \cite{vanderweele2009conceptual} for additional discussion of these assumptions.

\subsection{Direct and Indirect Effects}

We review the terms natural direct effect, natural indirect effect, and total effect as used in \cite{vanderweele2010odds} and \cite{tchetgen2012semiparametric}. These definitions admit a decomposition of total effect into indirect (effect of $X$ on $Y$ passing through $\bs{M}$) and direct (effect of $X$ on $Y$ not through $\bs{M}$) effects. We analyze these on the mean difference scale, the odds scale (useful with binary outcomes), and the restricted mean difference scale (useful with survival outcomes).

\subsubsection{Mean Difference Scale}
The average direct effect on the mean difference scale when changing $X_i$ from $x'$ to $x''$ with respect to mediators $\bs{M}$ is defined as
\begin{equation}
\label{eq:de}
DE_{X_i}(x',x'') = \E[Y^{X_i=x'',\bs{M}^{X_i=x'}} - Y^{X_i=x'}].
\end{equation}
The counterfactual random variable $Y^{X_i=x'',\bs{M}^{X_i=x'}}$ is the value $Y$ would have obtained had $X_i$ been set to the value $x''$ and $\bs{M}$ set to the value it would have obtained had $X_i$ been set to $x'$. In contrast $Y^{X_i=x'}$ is the value of $Y$ when $X_i$ is set to $x'$ (Note $Y^{X_i=x'} = Y^{X_i=x',\bs{M}^{X_i=x'}}$ by Assumption \ref{assump:8}). Thus the difference in these counterfactual quantities captures the intuitive notion of the change in $Y$ when the direct link from $X_i$ to $Y$ is changed from $x'$ to $x''$ but the indirect link (through $\bs{M}$) remains at $x'$.

The right hand side of Equation \ref{eq:de} cannot be directly estimated because it depends on counterfactual random variables which are not observed. However it is possible to express the direct effect as a function of the joint distribution of observed random variables which then facilitates estimation.

\begin{theorem}[Direct Effect Mean Difference]
\label{thm:direct_effect}
Assuming \ref{assump:1} -- \ref{assump:8}
\begin{align*}
DE_{X_i}(x',x'') &= \underbrace{\int \E[Y|x'',\bs{x}_{-i},\bs{m},\bs{c}]p(\bs{m}|x',\bs{x}_{-i},\bs{c})p(\bs{x}_{-i},\bs{c})d\bs{x}_{-i}d\bs{m}d\bs{c}}_{\equiv e(x',x'')},\\
&- \underbrace{\int\E[Y|x',\bs{x}_{-i},\bs{m},\bs{c}]p(\bs{m}|x',\bs{x}_{-i},\bs{c})p(\bs{x}_{-i},\bs{c})d\bs{x}_{-i}d\bs{m}d\bs{c}}_{\equiv e(x',x')}.
\end{align*}
\end{theorem}
See the Appendix for a proof of this result. Estimators of these quantities are discussed in Section \ref{sec:computation}. Similarly the natural indirect effect is defined as 
\begin{equation}
IE_{X_i}(x',x'') = \E[Y^{X_i=x''} - Y^{X_i=x'',\bs{M}^{X_i=x'}}].
\end{equation}
Again the natural indirect effect can be represented in terms of the joint probability distribution of the observed random variables.

\begin{theorem}[Indirect Effect Mean Difference]
Assuming \ref{assump:1} -- \ref{assump:8}
\begin{align*}
IE_{X_i}(x',x'') &= \underbrace{\int E[Y|x'',\bs{x}_{-i},\bs{m},\bs{c}]p(\bs{m}|x'',\bs{x}_{-i},\bs{c})p(\bs{x}_{-i},\bs{c})d\bs{x}_{-i}d\bs{m}d\bs{c}}_{\equiv e(x'',x'')},\\
&- \underbrace{\int E[Y|x'',\bs{x}_{-i},\bs{m},\bs{c}]p(\bs{m}|x',\bs{x}_{-i},\bs{c})p(\bs{x}_{-i},\bs{c})d\bs{x}_{-i}d\bs{m}d\bs{c}}_{\equiv e(x',x'')}.
\end{align*}
\end{theorem}
The proof follows similar reasoning to the proof of Theorem \ref{thm:direct_effect}. Finally we have the general mediation formula
\begin{equation*}
TE_{X_i}(x',x'') = DE_{X_i}(x',x'') + IE_{X_i}(x',x'') = e(x'',x'') - e(x',x').
\end{equation*}
The mediation formula states that the total effect is the sum of the direct and indirect effects. The relative contributions of direct and indirect effect are important for understanding the paths by which $X_i$ causes changes in $Y$. For example if there is no direct effect, then all changes in $Y$ caused by $X_i$ pass through $\bs{M}$. 

\subsubsection{Odds Scale}
The total, direct, and indirect effects require computing three quantities, $e(x'',x''),e(x',x''),e(x',x')$. For binary outcome $y$ \cite{vanderweele2010odds} defined the total, direct, and indirect effects on the odds scale:
\begin{align}
\label{eq:effects_odds}
TE^o(x',x'') &= \frac{\frac{e(x'',x'')}{1-e(x'',x'')}}{\frac{e(x',x')}{1-e(x',x')}}, \nonumber \\
DE^o(x',x'') &= \frac{\frac{e(x'',x'')}{1-e(x'',x'')}}{\frac{e(x',x'')}{1-e(x',x'')}}, \\
IE^o(x',x'') &= \frac{\frac{e(x',x'')}{1-e(x',x'')}}{\frac{e(x',x')}{1-e(x',x')}}. 
\end{align}
The effect decomposition is now
\begin{equation*}
TE^o(x',x'') = DE^o(x',x'') IE^o(x',x'').
\end{equation*}
%Such a decomposition is useful in case--control studies where the mean difference scale cannot be consistently estimated. See \cite{vanderweele2010odds} for a discussion. 

\subsubsection{Restricted Mean Difference Scale}
\label{sec:rms}

With survival outcomes, estimators of the expected response (i.e. $\E[Y|\bs{x},\bs{m},\bs{c}]$) often have high variance in the presence of censoring. Instead, we consider mean survival time restricted to a fixed time $L$ i.e. $E[\min(Y,L)|\bs{x},\bs{m},\bs{c}]$ \citep{chen2001causal}. The restricted mean is interpreted as population average of the amount of survival time experienced during the initial $L$ time of follow-up, providing an interpretable and clinically meaningful summary of the survival in the presence of censoring \citep{uno2014moving}. The definitions of direct, indirect, and total effects can be applied to the restricted mean survival scale.

The direct effect on the restricted mean scale is defined as
\begin{equation*}
DE^R_{X_i}(x',x'') = \E[\min(Y^{X_i=x'',\bs{M}^{X_i=x'}},L) - \min(Y^{X_i=x'},L)].
\end{equation*}
\begin{theorem}[Direct Effect Restricted Mean]
\label{thm:direct_rms}
Assuming \ref{assump:1} -- \ref{assump:8}
\begin{align*}
DE^R_{X_i}(x',x'') &= \underbrace{\int \E[min(Y,L)|x'',\bs{x}_{-i},\bs{m},\bs{c}]p(\bs{m}|x',\bs{x}_{-i},\bs{c})p(\bs{x}_{-i},\bs{c})d\bs{x}_{-i}d\bs{m}d\bs{c}}_{\equiv e^R(x',x'')},\\
&- \underbrace{\int \E[min(Y,L)|x',\bs{x}_{-i},\bs{m},\bs{c}]p(\bs{m}|x',\bs{x}_{-i},\bs{c})p(\bs{x}_{-i},\bs{c})d\bs{x}_{-i}d\bs{m}d\bs{c}}_{\equiv e^R(x',x')}.
\end{align*}
\end{theorem}
See the Appendix for a proof of this result. Similarly the natural indirect effect is defined as 
\begin{equation*}
IE^R_{X_i}(x',x'') = \E[min(Y^{X_i=x''},L) - min(Y^{X_i=x'',\bs{M}^{X_i=x'}},L)]
\end{equation*}
The natural indirect effect can be represented in terms of the joint probability distribution of the observed random variables.
\begin{theorem}[Indirect Effect Mean Difference]
Assuming \ref{assump:1} -- \ref{assump:8}
\begin{align*}
IE^R_{X_i}(x',x'') &= \underbrace{\int E[min(Y,L)|x'',\bs{x}_{-i},\bs{m},\bs{c}]p(\bs{m}|x'',\bs{x}_{-i},\bs{c})p(\bs{x}_{-i},\bs{c})d\bs{x}_{-i}d\bs{m}d\bs{c}}_{\equiv e^R(x'',x'')},\\
&- \underbrace{\int E[min(Y,L)|x'',\bs{x}_{-i},\bs{m},\bs{c}]p(\bs{m}|x',\bs{x}_{-i},\bs{c})p(\bs{x}_{-i},\bs{c})d\bs{x}_{-i}d\bs{m}d\bs{c}}_{= e^R(x',x'')}.
\end{align*}
\end{theorem}
The proof follows similar reasoning to the proof of Theorem \ref{thm:direct_rms}. The mediation formula again holds on the restricted mean scale
\begin{equation*}
TE^R_{X_i}(x',x'') = DE^R_{X_i}(x',x'') + IE^R_{X_i}(x',x'') = e^R(x'',x'') - e^R(x',x').
\end{equation*}

\subsection{Probability Models}
\label{sec:probability_models}
The probabilistic relationships among the variables in the Figure \ref{fig:dag} DAG are specified with parametric and semi--parametric statistical models for any configurations $\bs{c}$, $\bs{x}$, $\bs{m}$ and $y$ of $\bs{C}$, $\bs{X}$, $\bs{M}$ and $Y$, respectively. We assume linear relations for the conditional distribution of $\bs{M}$ given $\bs{X}$ and $\bs{C}$. Specifically,
\begin{equation}
\label{eq:mmodel}
\bs{m} = \bs{\beta}^{(X)}\bs{x} + \bs{\beta}^{(C)}\bs{c}+ \bs{\beta}^{(0)} + \bs{\epsilon},
\end{equation}
where $\bs{\beta}^{(X)} =(\beta_{j,i}^{(X)})\in \mathbb{R}^{r\times p}$, $\bs{\beta}^{(C)}=(\beta^{(C)}_{j,i})\in \mathbb{R}^{r \times q}$, $\bs{\beta}^{(0)}=(\beta^{(0)}_j) \in \mathbb{R}^r$, $\bs{\epsilon} \sim N_r(0,\bs{\Sigma}_\epsilon)$, and $\bs{\Sigma}_\epsilon \in \mathbb{R}^{r \times r}$ is a covariance matrix. In the case where mediators are conditionally independent given $\bs{X}$ and $\bs{C}$, $\bs{\Sigma}_\epsilon$ will be a diagonal matrix.

We consider three parametric models, linear, logistic, and Cox Proportional hazards, for linking $Y$ with $\bs{X}$, $\bs{M}$, and $\bs{C}$. Each of these models has parameters $\bs{\alpha} = (\bs{\alpha}^{(X)},\bs{\alpha}^{(M)},\bs{\alpha}^{(C)})$ where $\bs{\alpha}^{(X)} =(\alpha_j^{(X)})\in \mathbb{R}^p$, $\bs{\alpha}^{(M)}=(\alpha^{(M)}_j) \in \mathbb{R}^r$ and $\bs{\alpha}^{(C)}=(\alpha_j^{(C)}) \in \mathbb{R}^q$. The three models are:
\begin{itemize}
\item \underline{Linear:} 
\begin{equation}
\label{eq:ylmodel}
y = \bs{x}^T \bs{\alpha}^{(X)} + \bs{m}^T \bs{\alpha}^{(M)}  + \bs{c}^T \bs{\alpha}^{(C)} + \alpha^{(0)} + \delta,
\end{equation}
where $\delta \sim N(0,\sigma_\delta^2)$ independent of all other terms in the model and $\alpha^{(0)} \in \mathbb{R}^1$.
\item \underline{Logistic:}
\begin{equation}
\label{eq:ylogmodel}
Y \sim Bernoulli((1+e^{-(\bs{x}^T \bs{\alpha}^{(X)} + \bs{m}^T \bs{\alpha}^{(M)}  + \bs{c}^T \bs{\alpha}^{(C)} + \alpha^{(0)})})^{-1}),
\end{equation}
where $\alpha^{(0)} \in \mathbb{R}^1$.
\item \underline{Cox proportional hazards:} The failure time $Y$ is assumed to follow a hazard function model
\begin{equation}
\label{eq:ycoxmodel}
h(y|\bs{x},\bs{m},\bs{c}) = h_0(y)e^{\bs{x}^T \bs{\alpha}^{(X)} + \bs{m}^T \bs{\alpha}^{(m)}  + \bs{c}^T \bs{\alpha}^{(C)}},
\end{equation}
where $h_0$ is the unspecified baseline hazard.
\end{itemize}

\subsection{Estimation and Computation of Effects}
\label{sec:computation}

For linear models, the direct, indirect, and total effects have simple definitions in terms of path coefficients from the probability models in Section \ref{sec:probability_models}. For non--linear models, we estimate model coefficients and then numerically approximate indirect and direct effect integrals.

\subsubsection{Mean Difference and Odds Scale}

Both the mean difference and odds scale require estimates of three quantities: $e(x'',x''),e(x',x''),e(x',x')$. We discuss estimation of $e(x',x'')$. The algorithms for $e(x'',x'')$ and $e(x',x')$ are nearly identical. Recall
\begin{equation*}
e(x',x'') \equiv \int E[Y|X_i=x'',\bs{x}_{-i},\bs{m},\bs{c}]p(\bs{m}|X_i=x',\bs{x}_{-i},\bs{c})p(\bs{x}_{-i},\bs{c})d\bs{x}_{-i}d\bs{m}d\bs{c}.
\end{equation*}
We plug estimates into unknown quantities in the integrand and use Monte Carlo sampling to approximate the integral. The quantity $p(\bs{m}|X_i=x',\bs{x}_{-i},\bs{c})p(\bs{x}_{-i},\bs{c})$ is a distribution on $\bs{m},\bs{x}_{-i},\bs{c}$. We use the observed data samples $\bs{x}_{-i,l}$ and $\bs{c}_l$ for $l=1,\ldots,n$ as a draw from $p(\bs{x}_{-i},\bs{c})$. We then draw $\bar{\bs{m}}_l \sim \widehat{p}(\bs{m}|X_i=x',\bs{x}_{-i,l},\bs{c}_l)$. The bar in $\bar{\bs{m}}_l$ denotes the fact that this is data we simulate, not the actual observed mediator for sample $l$. The Monte Carlo approximation to the integral is
\begin{equation*}
\widehat{e}(x',x'') = \frac{1}{n}\sum_{l=1}^n \widehat{E}[Y|X_i = x'',\bs{x}_{-i,l},\bar{\bs{m}}_l,\bs{c}_l].
\end{equation*}
We specify estimates $\widehat{E}[Y|\bs{x},\bs{m},\bs{c}]$ using response models in Equations \eqref{eq:ylmodel} and \eqref{eq:ylogmodel}.
\begin{itemize}
\item \underline{Linear Model:}
\begin{equation*}
\widehat{E}[Y|\bs{x},\bs{m},\bs{c}] = \bs{x}^T \widehat{\bs{\alpha}}^{(X)} + \bs{m}^T \widehat{\bs{\alpha}}^{(M)}  + \bs{c}^T \widehat{\bs{\alpha}}^{(C)} + \widehat{\alpha}^{(0)},
\end{equation*}
\item \underline{Logistic Model:}
\begin{equation*}
\widehat{E}[Y|\bs{x},\bs{m},\bs{c}] = \widehat{p}(Y=1|\bs{x},\bs{m},\bs{c}) = \frac{1}{1 + e^{-\bs{x}^T \widehat{\bs{\alpha}}^{(X)} - \bs{m}^T \widehat{\bs{\alpha}}^{(M)}  - \bs{c}^T \widehat{\bs{\alpha}}^{(C)} - \widehat{\alpha}^{(0)}}}.
\end{equation*}
\end{itemize}
For $\widehat{p}(\bs{m}|\bs{x},\bs{c})$, recall by Equation \eqref{eq:mmodel} that 
\begin{equation*}
\bs{m} | \bs{x},\bs{c} \sim N(\bs{\beta}^{(X)}\bs{x} + \bs{\beta}^{(C)}\bs{c} + \bs{\beta}^{(0)},\bs{\Sigma}_\epsilon).
\end{equation*}
The $\bar{\bs{m}}_l$ are simulated from the plug--in based measure $\widehat{p}(\bs{m}|X_i=x',\bs{x}_{-i,l},\bs{c}_l)$. We estimate $\bs{\Sigma}_\epsilon$ using the sample covariance of the regression residuals $\bs{r}_l = \bs{m_l} - (\widehat{\bs{\beta}}^{(X)}\bs{x_l} + \widehat{\bs{\beta}}^{(C)}\bs{c_l} + \widehat{\bs{\beta}}^{(0)})$. If one makes the assumption of conditionally independent mediators given $\bs{x}$ and $\bs{c}$, i.e. $m_j \indep m_k | \bs{x},\bs{c}$, for all $j,k$ then $\bs{\Sigma}_\epsilon$ is diagonal and can be estimated using the error variances from separate univariate regressions, $m_j | (x,c)$ for all $j$. This estimate will be more efficient, but carries more assumptions.

\subsubsection{Restricted Mean Scale}

On the restricted mean scale, the quantities of interest are $e^R(x'',x''),e^R(x',x''),e^R(x',x')$. These are nearly identical to the terms for mean difference and odds scales with the exception that $Y$ is replaced by $min(Y,L)$ within the expectation. Thus the numerical approximation to the integral follows the procedure in Section 4.1. The numerical approximation to the integral can be accomplished by deriving estimates for the survival function $S(y|\bs{x},\bs{m},\bs{c}) = P(Y> y|\bs{x},\bs{m},\bs{c})$. An estimator for the restricted mean is 
\[\hat{\E}[min(Y,L)|\bs{x},\bs{m},\bs{c}] = \int_{0}^L \hat{S}(y|\bs{x},\bs{m},\bs{c})dy,\]
with estimates from the Cox proportional hazards model in equation (\ref{eq:ycoxmodel})
\begin{equation*}
\widehat{S}(y|\bs{x},\bs{m},\bs{c}) = e^{\left(-\int_0^y \widehat{h}_0(t)dt\right) e^{\bs{x}^T \widehat{\bs{\alpha}}^{(X)} + \bs{m}^T \widehat{\bs{\alpha}}^{(M)}  + \bs{c}^T \widehat{\bs{\alpha}}^{(C)}}},
\end{equation*}
where $\widehat{h}$ is an estimate of the baseline hazard function and $\widehat{\bs{\alpha}}^{(X)},\widehat{\bs{\alpha}}^{(M)},\widehat{\bs{\alpha}}^{(C)}$ are coefficient estimates.

\subsection{Bootstrap Based Confidence Intervals and Hypothesis Tests}

There are several existing approaches for creating confidence intervals and performing hypothesis tests in mediation analysis. The problem of hypothesis testing for the existence of an indirect effect has generated particular interest because it is practically important and challenging, due to the composite nature of the null hypothesis \citep{barfield2017testing}. In univariate linear models, the null hypothesis of no indirect effect is $H_0: \beta^{(X)}\alpha^{(M)} = 0$. Thus the null can be true if either there is no exposure-mediator causal effect or if there is no mediator-response causal effect. Delta method based approximations to the sampling distribution are not valid due to the non-normality of $\widehat{\beta}^{(X)}\widehat{\alpha}^{(M)}$ under the null hypothesis. The joint significance test proposes computing p-values for the tests $H_0: \beta^{(X)}=0$ and $H_0:\alpha^{(M)} = 0$. The maximum of these p-values controls Type I error. This control is conservative in the case where both the exposure--mediator and mediator--response relations are null, i.e. $\beta^{(x)} = \alpha^{(m)} = 0$.

We propose computing confidence intervals and hypothesis tests using bootstrap sampling quantiles. Suppose $B$ bootstrap samples of the data are taken. Let $\widehat{IE}_{X_i}^{(b)}(x',x'')$ be the estimated indirect effect when changing $x_i$ from $x'$ to $x''$ in bootstrap sample $b=1,\ldots,B$. Then a $(1-\alpha)100$\% confidence interval for $IE_{X_i}(x',x'')$ has endpoints at the $\alpha/2$ and $1-\alpha/2$ quantiles of the $\widehat{IE}_{X_i}^{(b)}(x',x'')$ distribution. For testing the hypothesis
\begin{align}
\label{eq:test}
&IE_{X_i}(x',x'') = \Delta \\
&IE_{X_i}(x',x'') \neq \Delta, \nonumber
\end{align}
let $p_L$ and $p_U$ be the proportion of bootstrap samples below and above $\Delta$, respectively. Specifically $p_L = B^{-1} \sum_{b=1}^B \ind{\widehat{IE}_{X_i}^{(b)}(x',x'') < \Delta}$ and $p_U = B^{-1} \sum_{b=1}^B \ind{\widehat{IE}_{X_i}^{(b)}(x',x'') > \Delta}$. Then the p-value for hypothesis test \eqref{eq:test} is $2\min(p_L,p_U)$. Similar procedures can be used to construct confidence intervals and test for direct effects. Following \cite{efron1994introduction} Chapter 13, we compute $B=1000$ bootstrap samples for making confidence intervals. Larger numbers of bootstrap samples could be used to ensure that the quantiles of the bootstrap samples better approximate the bootstrap sampling distribution, at the cost of additional computation time.

\subsection{Relation to Existing Work}
\label{sec:prior_work}

In our framework direct and indirect effects are estimated by approximating integrals. Under additional assumptions on the joint distribution of random variables, direct and indirect effects are approximately simple functions of coefficients. For example with the logistic model with univariate mediator if $P(y=1) \approx 0$ then $DE^o(x',x'+1) \approx \exp(\alpha^{(X)})$ and $IE^o(x',x'+1) \approx \exp(\beta^{(X)}\alpha^{(M)})$. One can then estimate these approximations via logistic regression estimates of $\alpha^{(X)}$ and $\beta^{(X)}$. This estimator is increasingly accurate as the disease becomes more rare, i.e. $P(y=1)$ converges to $0$ \citep{huang2014joint,vanderweele2010odds}. Along the same line, \cite{gaynor2018mediation} proposed a probit approximation to the logistic function, designed for common responses where the rare disease assumption does not hold. Our framework avoids these additional assumptions by directly approximately the direct effect and indirect effect integrals at the cost of increased computation time.

\cite{fasanelli2019marginal} proposed mediation analysis for survival outcomes through specifications of a response model $p(y|x_i,\bs{c},\bs{m})$ and a propensity model $p(x_i|\bs{c})$, using inverse probability weighting to estimate the causal effects. The approach avoids specification of a model for mediators but can only accommodate a binary or categorical exposure $X$. Thus it could not be applied to the data example in Section \ref{sec:application} which considers continuous gene expressions as exposures.

Our computational framework of approximating integrals is closest to that proposed in \cite{imai2010general}. However we offer additional functionality by 1) incorporating multiple mediators that may form a correlation structure without having to specify any internal mediator causal structure and 2) modelling effects on the odds and restricted mean scale, which are more appropriate than the mean difference for the logistic and Cox proportional hazards models.

\section{Simulations}
\label{sec:sim}

\subsection{Logistic Models: Bias}
\label{sec:logistic_sim}

\begin{figure}[H]
  \centering
a) \includegraphics[width=0.45\textwidth]{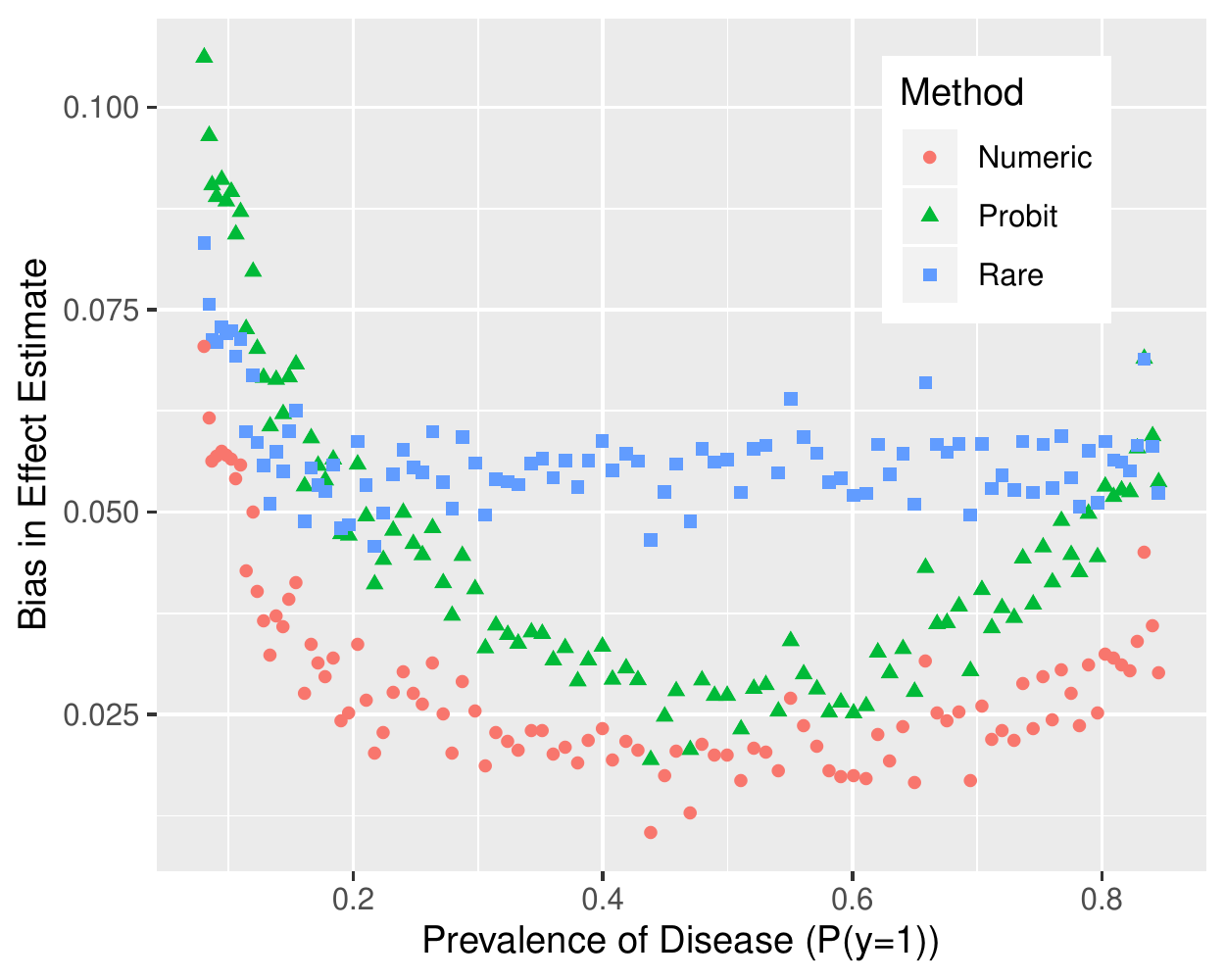}
b) \includegraphics[width=0.45\textwidth]{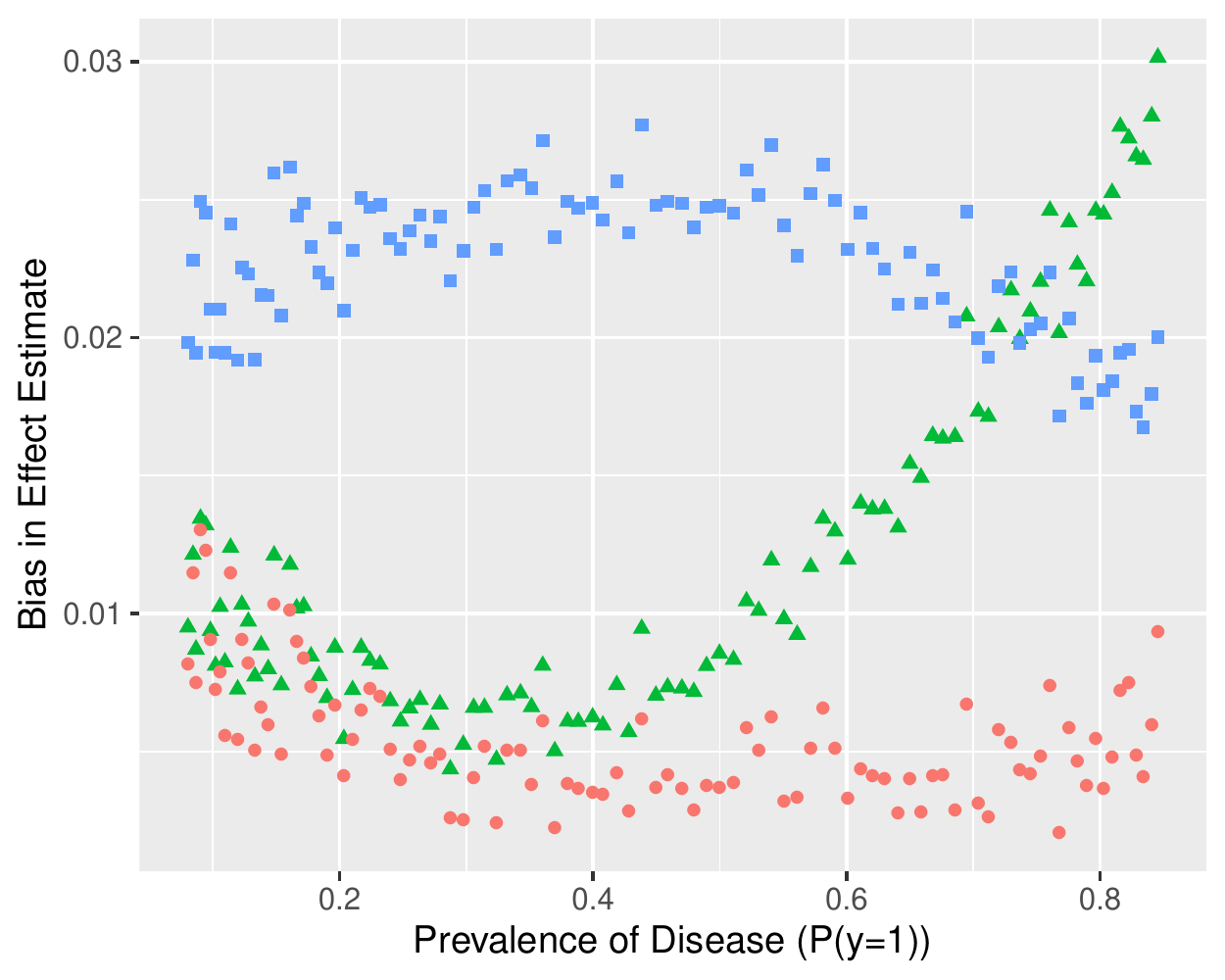}
\caption{Comparison of methods for computing the a) direct effect and b) indirect effect with logistic models. Numeric approximation has lower bias than the rare disease approximation and the probit approximation.\label{fig:logistic_approx}}
\end{figure}

We compare the performance of our method with two approximation methods that exploit rare disease assumption \citep{vanderweele2010odds} and probit model \citep{gaynor2018mediation} in the context of binary outcomes. We focus on direct and indirect effect estimator bias as a function of disease prevalence.

Following \cite{gaynor2018mediation} (Section 3.1) we simulate
\begin{align*}
c &\sim N(0.12,0.75^2)\\
x &\sim N(0.4,0.75^2)\\
m|x,c &\sim N(0.1 + 0.5x + 0.4c,0.75^2)\\
logit(P(y=1|x,m,c)) &= k + 0.4x + 0.5m + 0.25c.
\end{align*}
The constant $k$ is varied to generate different prevalences (different values of $p(y=1)$). The sample size of $n=500$ is generated $N=5000$ times. For each run, the three estimators are computed. The estimators are averaged across the runs and the bias of the estimator is computed. The bias as a function of prevalence is shown in Figure \ref{fig:logistic_approx}. The scatter in the points is due to Monte Carlo approximation on (setting $N=\infty$ would result in smooth curves). Our estimation strategy has lower bias than the methods based on the rare disease assumption and probit approximations at all prevalences for both the direct and indirect effect.

\subsection{Survival Outcomes: Type I Error Control}

\begin{figure}[H]
  \centering
\includegraphics[width=0.6\textwidth]{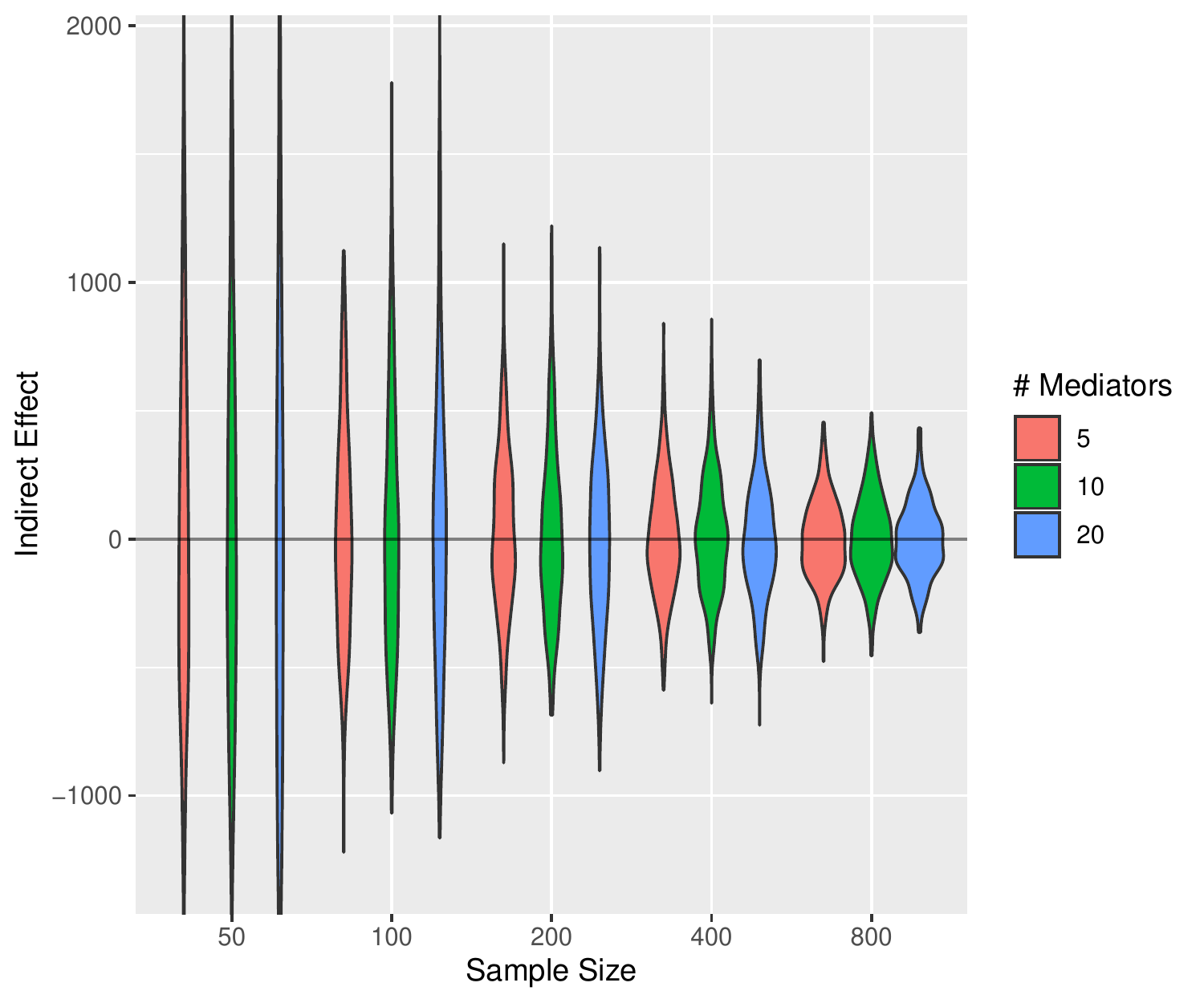}
\caption{Indirect effect point estimates for 500 runs with 0 indirect effect. \label{fig:sim_null_indirect_0}}
\end{figure}

We study dependence of Type I Error on the number of candidate mediators and sample size when the true indirect effect is $0$ using survival responses. We simulate a single causal binary exposure $X \in \mathbb{R}^1$ with prevalence probability 0.5 and 5, 10, and 20 candidate mediators. 

Five candidate mediators are generated as linear functions of exposure with $R^2 = 0.2$. The remaining candidate mediators (0, 5, or 15 for the simulations with a total of 5, 10, and 20 total candidate mediators respectively) are uncorrelated with exposure. The response follows an exponential model with Cox proportional hazards model coefficient $0.5$ for the exposure direct effect with 50\% censoring. The candidate mediators have coefficients of $0$ in the response model (hazard function depends on exposure only), so the true indirect effect is $0$. At sample sizes of 50, 100, 200, 400, and 800, we simulate 500 runs and compute the indirect effect estimate. Violin plots of the results are shown in Figure \ref{fig:sim_null_indirect_0}. The estimates are somewhat biased positive with both bias and variance decreasing as the sample size increases.

We use the bootstrap quantile method to compute 95\% confidence intervals. The results are shown in Table \ref{tab:sim_null}. Overall, the confidence intervals have coverage probability near or above the nominal level. When the number of mediators are relatively large compared to the sample size, as expected, the coverage probability was increased larger than 0.95 due to large confidence intervals resulting from unstable parameter estimates across bootstrap samples. 
\begin{center}
% latex table generated in R 3.5.1 by xtable 1.8-3 package
% Mon Jan  6 09:34:01 2020
\begin{table}[ht]
\centering
\begin{tabular}{c|c|c}
 n & No. Med. & CI Cov. \\ 
  \hline
50 & 5 & 0.97 \\ 
   \hline
50 & 10 & 0.98 \\ 
   \hline
50 & 20 & 1.00 \\ 
   \hline
100 & 5 & 0.97 \\ 
   \hline
100 & 10 & 0.97 \\ 
   \hline
100 & 20 & 0.99 \\ 
   \hline
200 & 5 & 0.96 \\ 
   \hline
200 & 10 & 0.95 \\ 
   \hline
200 & 20 & 0.97 \\ 
   \hline
400 & 5 & 0.94 \\ 
   \hline
400 & 10 & 0.95 \\ 
   \hline
400 & 20 & 0.96 \\ 
   \hline
800 & 5 & 0.94 \\ 
   \hline
800 & 10 & 0.94 \\ 
   \hline
800 & 20 & 0.96 \\ 
   \hline
\end{tabular}
\caption{Empirical coverage probabilities for 95\% confidence intervals in the null simulation (indirect effect=0).\label{tab:sim_null}} 
\end{table}

\end{center}

\subsection{Survival Outcomes: Power}
\label{sec:surv_power}
We simulate a single binary exposure $X$ with prevalence probability 0.5. We simulated 5 mediators with linear exposure-mediator relationships with $R^2$ of $0.2$. The response $Y$ is again generated from an exponential model with approximately 50\% censoring. We consider two cases: strong mediators with mediator-response path coefficients of $0.2$ and weak mediators with mediator-response path coefficients of $0.1$.

\begin{figure}[H]
  \centering
a) \includegraphics[width=0.45\textwidth]{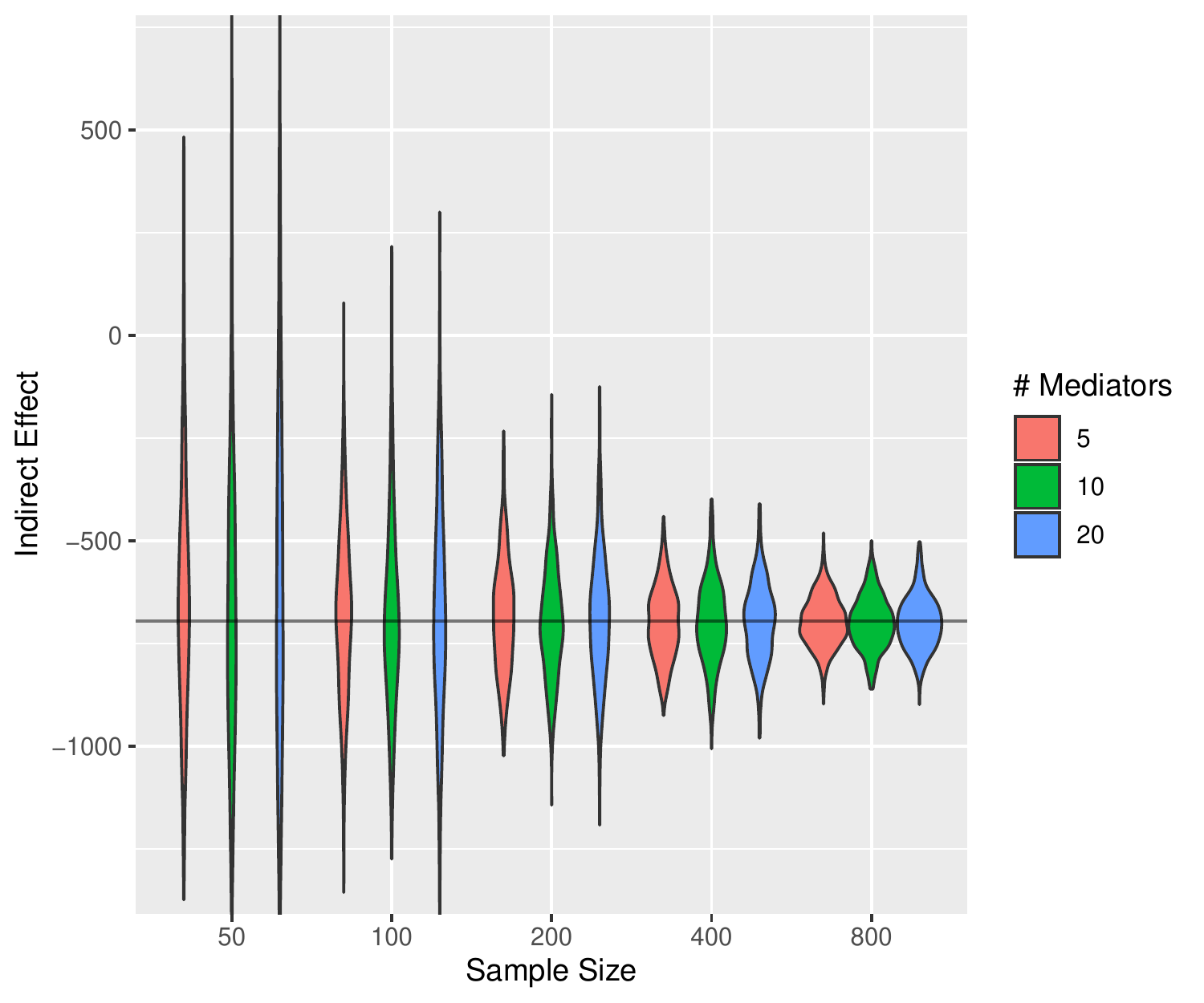}
b) \includegraphics[width=0.45\textwidth]{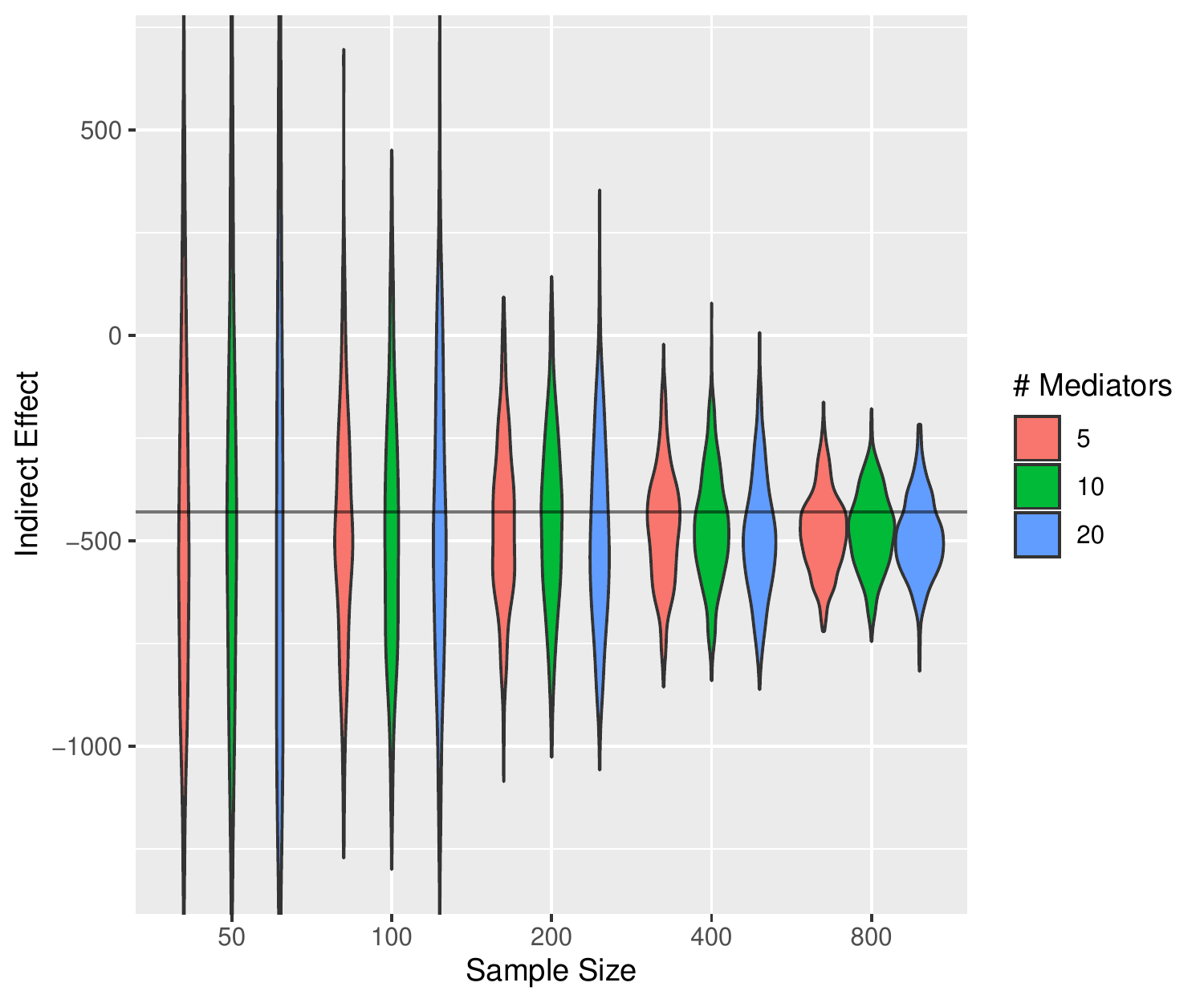}
\caption{a) Indirect effect point estimates for 500 runs with large indirect effect. Horizontal line is true IE. b) Indirect effect point estimates for 500 runs with small indirect effect. Horizontal line is true IE.  \label{fig:sim_indep_results}}
\end{figure}

In order to evaluate the potential impact of false candidate mediators, we simulate with total number of candidate mediators, 5, 10, and 20, where 5 of the mediators are true mediators among the candidates and the remaining mediators are noise. Thus there are 0, 5, and 15 candidate mediators which are not true mediators. We considered sample sizes of 50, 100, 200, 400, and 800. The results are summarized over 500 simulation runs for each scenario. Violin plots of the results with strong mediators (mediator-response path coefficients are 0.2) are shown in Figure \ref{fig:sim_indep_results} a). Presence of noise mediators creates a noticeable bias at sample size of 50 and inflates the variance of the estimate (relative to 5 and 10 mediators) at sample sizes of 50 and 100. Empirically, the point estimates appear to be converging to the true indirect effect of -695.

With weak mediators (mediator-response path coefficients are 0.1), the true indirect effect is -429. The results are summarized in Figure \ref{fig:sim_indep_results} b). Broadly we see the same pattern as before. At sample sizes of 50, 100, and 200, the point estimates are often near 0, suggesting low power to conclude presence of an indirect effect.

Table \ref{tab:sim_indep} summarizes the power (test with Type I Error $\alpha=0.05$) and coverage probability (95\% confidence interval). We see favorable results with power at or near 100\% for sample sizes 200 and above with large indirect effects and for sample size of 400 and 800 with small indirect effects. Since the real data in Section \ref{sec:application} has 470 observations, this gives us confidence that we are able to detect even weak mediators with reasonable power.

\begin{center}
% latex table generated in R 3.5.1 by xtable 1.8-3 package
% Mon Jan  6 09:33:44 2020
\begin{table}[ht]
\centering
\begin{tabular}{c|c|cc|cc}
   & & \multicolumn{2}{c}{Strong Med.} & \multicolumn{2}{c}{Weak Med.} \\
 \hline
n & No. Med. & CI Cov. & Power & CI Cov. & Power \\ 
  \hline
50 & 5 & 0.99 & 0.34 & 0.98 & 0.13 \\ 
   \hline
50 & 10 & 0.99 & 0.15 & 0.99 & 0.06 \\ 
   \hline
50 & 20 & 0.98 & 0.00 & 1.00 & 0.00 \\ 
   \hline
100 & 5 & 0.97 & 0.79 & 0.96 & 0.31 \\ 
   \hline
100 & 10 & 0.97 & 0.68 & 0.97 & 0.23 \\ 
   \hline
100 & 20 & 0.99 & 0.41 & 0.99 & 0.10 \\ 
   \hline
200 & 5 & 0.96 & 0.98 & 0.95 & 0.56 \\ 
   \hline
200 & 10 & 0.97 & 0.98 & 0.95 & 0.47 \\ 
   \hline
200 & 20 & 0.98 & 0.93 & 0.96 & 0.42 \\ 
   \hline
400 & 5 & 0.98 & 1.00 & 0.95 & 0.87 \\ 
   \hline
400 & 10 & 0.94 & 1.00 & 0.94 & 0.83 \\ 
   \hline
400 & 20 & 0.98 & 1.00 & 0.94 & 0.82 \\ 
   \hline
800 & 5 & 0.97 & 1.00 & 0.92 & 0.99 \\ 
   \hline
800 & 10 & 0.96 & 1.00 & 0.93 & 0.99 \\ 
   \hline
800 & 20 & 0.97 & 1.00 & 0.94 & 0.99 \\ 
   \hline
\end{tabular}
\caption{Empirical coverage probabilities and power for simulation with strong mediators and weak mediators.\label{tab:sim_indep}} 
\end{table}

\end{center}

\section{Data Application}
\label{sec:application}
The Cancer Genome Atlas (TCGA) project collected and studied genetic changes in cancer patients at the genomic, transcriptomic, and proteomic levels. This comprehensive multi-omic data set enables modeling dependencies across multiple platforms as well as associations with clinical variables such as patients' survival times. Amongst many other discoveries, \cite{cancer2013comprehensive} identified five core metabolic pathways in Kidney Clear Cell Carcinoma (see Figure S59), comprised of mRNAs and proteins which were associated with aggressive cancers. Kidney clear cell carcinoma has increasingly been identified as a metabolic disease and metabolic pathways are considered to be therapeutic targets of intervention \citep{rathmell2018metabolic}. Here we assess whether the causal effect of changes in these metabolic pathways at the mRNA level is mediated by changes at the metabolic protein expression level. Since increased mRNA expression levels have the ability to increase protein expression levels via translational mechanisms, it is sensible to view metabolic protein expression levels as potential causal mediators of the mRNA--survival relations.

%These pathways and the associated mRNA are listed in Table \ref{tab:tcga_paths} 
 
We include 470 patients with mRNA, protein, and survival data available. Each pathway is summarized at the mRNA level by computing the standardized first principal component (PC) for genes within the pathway. Larger component scores indicate higher expression in the pathway. The five metabolic proteins from Figure S59 of \cite{cancer2013comprehensive} (which are correlated with survival and in the same pathways as the mRNA) are treated as potential mediators: AMPKA alpha, AMPK pT172, ACC pS79, ACC, and PTEN. In the context of the DAG of Figure \ref{fig:dag}, the 5 mRNA pathway scores are the second layer (exposures), the 5 proteins are the third layer (mediators), and survival is the fourth layer (response). The median follow--up time is 1731 days ([1525,1871] 95\% CI) and the median survival time is 2564 days ([2190,$\infty$) 95\% CI) with 165 deaths observed out of 470 patients. 

\begin{figure}[H]
  \centering
\includegraphics[width=1\textwidth]{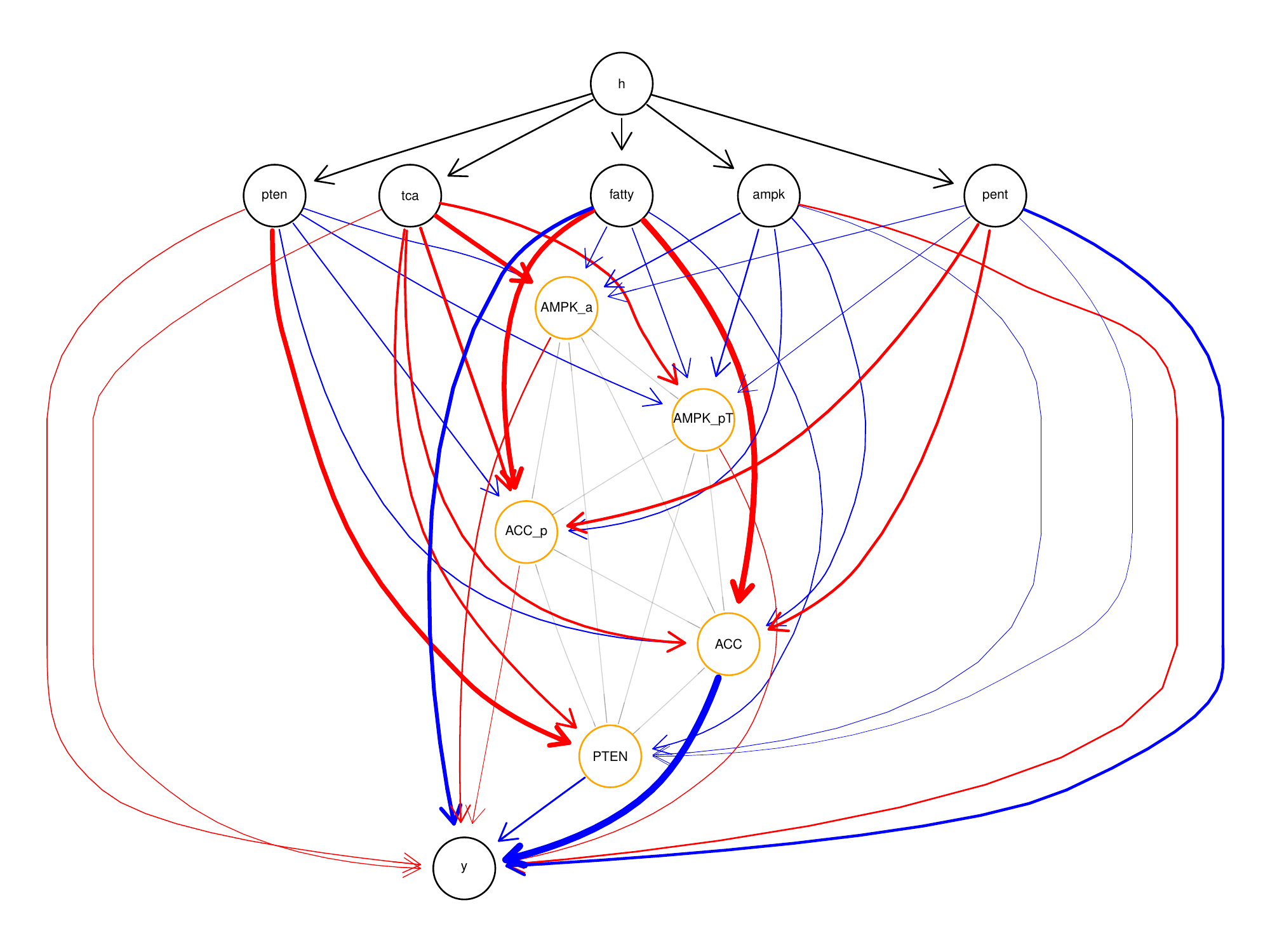}
  \caption{Graph illustrating causal assumptions and parameter coefficient estimates. Black nodes are mRNA pathway scores, orange nodes are protein expressions, and $y$ is survival. Edges are colored red for positive correlation and blue for negative correlation with edge width proportional to estimated coefficient value. \label{fig:TCGA_dag}}
\end{figure}

Figure \ref{fig:TCGA_dag} illustrates the estimated coefficients from the mediation model in Equation \eqref{eq:mmodel} and outcome model in Equation \eqref{eq:ycoxmodel} with black and orange nodes for mRNA pathways and proteins and the outcome node, and edges weighted and colored by the estimates.  The causal structure among the proteins is left unspecified by the model and is represented by undirected grey arrows between each pair of proteins. Edges are colored red for positive correlation and blue for negative correlation with edge width proportional to the absolute size of the coefficient estimate.
 
Using the methodology proposed in this work, we compute direct, indirect, and total effects for each pathway score using restricted mean survival truncated to $2000$ days and letting $x'$ and $x''$ be the 2.5 and 97.5 percentiles of each pathway score. The percentiles were chosen to represent a large change in pathway score still within the range of the observed data. Table \ref{tab:tcga_results} contains direct, indirect, and total effects as well as 95\% confidence intervals based on $B=1000$ bootstrap samples with responses measured in days. The TCA cycle, Pentose phosphate, and Fatty acid synthesis pathways have significant total effects at level $\alpha=0.05$ in the same directions found in \cite{cancer2013comprehensive}. On average, patients with high gene expression in the TCA cycle pathway live 329 days longer than the low expression group during 2000 days of follow-up. In contrast, patients with high gene expression in the Pentose phosphate pathways live, on average, 274 days shorter than the low expression groups during 2000 days of follow-up. TCA cycle and Pentose phosphate effects appear to be primarily direct. Fatty acid synthesis (FAS) has the largest absolute indirect effect point estimate. A FAS score change from the 2.5 to the 97.5 percentile has a total effect of reducing mean restricted lifetime by 446 days (1.22 years) with 156 days explained by changes in metabolic protein mediators, 35\% (156/446) of the total effect.

%of -156 days, representing 35\% (156/446) of the total effect. \mj{We need detailed interpretation  (1) explain the range of the FAS score and 2.5 and 97.5 percentiles; (2) Interpret the total and indirect effect in the restricted mean scale. For example, the restricted mean lifetime up to 2000 days (5.48 years) for KIRC patients who had xxx (the 97.5 percentile) of the FAS score is estimated to be 446 days (1.22 years) lower than the patients who had xxx (2.5 percentile). 35\% (156 days) of the total effect of 446 days are explained by the metabolic protein mediators. ]} 
The model coefficients (see Figure \ref{fig:TCGA_dag}) suggest that the indirect effect for FAS is primarily through the ACC protein. Specifically FAS is positively correlated with ACC which is positively correlated with survival. This is the same direction as the direct effect of FAS (blue line between fatty and y nodes). This observation is consistent with existing experimental data showing that FAS is mainly regulated via phosphorylation and dephosphorylation of ACC proteins \citep{kim1989role,davis2000overproduction,hardie1989regulation}.

% latex table generated in R 3.5.1 by xtable 1.8-3 package
% Mon Aug 24 15:22:26 2020
\begin{table}[ht]
\centering
\begin{tabular}{lrlrlrl}
  \hline
  Pathway & \multicolumn{2}{c}{Indirect} & \multicolumn{2}{c}{Direct}  & \multicolumn{2}{c}{Total} \\
 \hline
PTEN & -29 & [-146,77] & 203 & [-49,409] & 174 & [-74,384] \\ 
  TCA cycle & 40 & [-23,120] & 289 & [42,494] & 329 & [117,525] \\ 
  Fatty acid synthesis & -156 & [-339,20] & -290 & [-537,-91] & -446 & [-654,-268] \\ 
  AMPK & 23 & [-74,126] & 8 & [-292,316] & 30 & [-245,328] \\ 
  Pentose phosphate & -94 & [-247,79] & -181 & [-511,91] & -274 & [-574,-29] \\ 
   \hline
\end{tabular}
\caption{Indirect, Direct, and Total effects and 95\% confidence intervals (in days) of metabolomic mRNA expression as mediated by protein expression.\label{tab:tcga_results}} 
\end{table}

We further investigated mediation effects of non-metabolic proteins by considering 12 additional protein pathways representing biological functions such as apoptosis, DNA repair, and epithelial--mesanchymal transition. These pathways were previously studied for their roles in tumor cell behavior and therapy response \citep{ha2018personalized,bhattacharyya2020personalized,akbani2014pan}. Results are contained in Supplementary Tables S1--S12. We found evidence to suggest that TCA cycle is mediated by several protein pathways including Core reactive, TSC/mTOR, RAS/MAPK, and PI3K/AKT (p--value < 0.05).

\section{Discussion}
\label{sec:discussion}
We proposed a general and unified methodology of mediation analysis for data observed from random variables that form a multi-layered graphical structure. Direct and indirect effects are easily computed from standard probability models for different choices of outcomes such as continuous, binary and survival, and measured on the mean, odds and restricted mean scales from linear, logistic and Cox-proportional hazards models, respectively. The proposed framework has advantages over existing approaches such as not requiring assumptions on disease prevalence (rare or common disease assumptions) in the case of binary outcomes and accommodating continuous exposure variables ($x$) such as mRNA expression. The framework controls for confounders and accommodates correlated mediators without requiring assumptions on the direction of any mediator causal structure. Our \texttt{mediateR} package makes these models easily accessible to users.

Standard statistical tools (e.g. confidence intervals and hypothesis tests) can be used to assess the existence and likely ranges for the direct and indirect effects. Simulation studies with our method suggest that for small numbers of exposures and confounders ($<20$), the models produce reasonable parameter estimates and well calibrated uncertainties when samples sizes are in the hundreds. In high dimensional settings, regularization penalties could be used in the model fitting steps to estimate parameters. The direct and indirect effect integral approximations proposed in Section \ref{sec:computation} could then be used with these regularized parameter estimates.

The causal interpretation of direct and indirect effects requires strong causal assumptions (\ref{assump:1} -- \ref{assump:8}). These causal assumptions include no unmeasured confounders and no variable measurement error. These causal assumptions could be violated in a number of ways. For example, in the context of Figure \ref{fig:TCGA_dag}, a transcription factor protein not included in the model could have a causal effect on both Fatty Acid Synthesis (FAS, node name fatty) and independently on the ACC protein. To a limited extent these assumptions can be checked, and violations addressed, with additional modeling. For example, sensitivity analyses can be used to test for unobserved pre-exposure covariates \citep{imai2010identification}. Mediator measurement error which biases effect size estimates and can be corrected via regression calibration \citep{valeri2014mediation}. 

%\backmatter

\section*{Acknowledgments}
KAD was partially supported by the National Institutes of Health: P30CA016672, SPORE P50CA140388, CCTS TR000371, and by CPRIT RP160693. MJH was partially supported by the NIH/NCI 5R21CA220299. JPL was partially supported by National Institutes of Health SPORE P50CA127001 and SPORE P50CA140388 and CPRIT RP160693.  The authors acknowledge the support of the High Performance Computing facility at the University of Texas MD Anderson Cancer Center for providing computational resources (including consulting services) that have contributed to the research results reported in this paper. 

% \subsection*{Author contributions}
% 
% This is an author contribution text. This is an author contribution text. This is an author contribution text. This is an author contribution text. This is an author contribution text. 
% 
% \subsection*{Financial disclosure}
% 
% None reported.
% 
% \subsection*{Conflict of interest}
% 
% The authors declare no potential conflict of interests.

\section*{Data Availability and Supporting Information}

\textbf{Data:} The results here are in part based upon data generated by the TCGA Research Network \url{https://www.cancer.gov/tcga}.

\noindent
\textbf{\texttt{R} code for reproducing results:} Code for fitting these models is available in the R package \texttt{mediateR} on github \url{https://github.com/longjp/mediateR}.
%The \texttt{mediateR} package, written in the R programming language \citep{R}, uses the R--packages \texttt{glmnet} \citep{glmnet,coxnet}, \texttt{MASS} \citep{MASS}, \texttt{survival} \citep{survival-book}, \texttt{graph} \citep{graph-pack}, \texttt{knitr} \citep{knitr}, \texttt{rmarkdown} \citep{rmarkdown},  and \texttt{kableExtra} \citep{kableExtra}.
Code for reproducing all results in this work is available on github \url{https://github.com/longjp/mediateR_paper}. This code includes script based downloads of TCGA data using TCGA Assembler \citep{wei2017tcgawei2017tcga}.

\noindent
\textbf{Tables S1--S12.}
Mediation effects for non--metabolic protein pathways.

\appendix

\subsection*{Proof of Theorems \ref{thm:direct_effect} and \ref{thm:direct_rms}}
\label{sec:direct_proof}

Let $g: \mathbb{R}^1 \rightarrow \mathbb{R}^1$. In Theorem \ref{thm:direct_effect}, $g(\cdot) = \cdot$, and in Theorem \ref{thm:direct_rms}, $g(\cdot) = \min(\cdot,L)$. We have
\begin{align}
\E[g(Y^{X_i=x'',\bs{M}^{X_i=x'}})] &= \int_y g(y) p(Y^{X_i=x'',\bs{M}^{X_i=x'}}=y) \nonumber \\
&= \int_{y,\bs{x}_{-i},\bs{m},\bs{c}} g(y) \underbrace{p(Y^{X_i=x'',\bs{M}^{X_i=x'}}=y|\bs{x}_{-i},\bs{M}^{X_i=x'}=\bs{m},\bs{c})}_{\equiv A}\underbrace{p(\bs{x}_{-i},\bs{M}^{X_i=x'}=\bs{m},\bs{c})}_{\equiv B}. \label{eq:temp1}
\end{align}
Roman numerals above equals signs reference the counterfactual assumption used in the paper. We have
\begin{align*}
B &= p(\bs{M}^{x_i=x'}=\bs{m}|\bs{x}_{-i},\bs{c})p(\bs{x}_{-i},\bs{c})\\
&\stackrel{\text{\ref{assump:1}}}{=} p(\bs{M}^{x_i=x'}=\bs{m}|X_i=x',\bs{x}_{-i},\bs{c})p(\bs{x}_{-i},\bs{c})\\
&\stackrel{\text{\ref{assump:2}}}{=} p(\bs{M}=\bs{m}|X_i=x',\bs{x}_{-i},\bs{c})p(\bs{x}_{-i},\bs{c})\\
&= p(\bs{m}|x',\bs{x}_{-i},\bs{c})p(\bs{x}_{-i},\bs{c}).
\end{align*}
Next we have
\begin{align*}
A &= p(Y^{X_i=x'',\bs{m}}=y|\bs{x}_{-i},\bs{M}^{X_i=x'}=\bs{m},\bs{c})\\
&\stackrel{\text{\ref{assump:4}}}{=} p(Y^{X_i=x'',\bs{m}}=y|\bs{x}_{-i},\bs{c})\\
&\stackrel{\text{\ref{assump:5}}}{=} p(Y^{X_i=x'',\bs{m}}=y|X_i=x'',\bs{x}_{-i},\bs{c})\\
&\stackrel{\text{\ref{assump:6}}}{=} p(Y^{X_i=x'',\bs{m}}=y|X_i=x'',\bs{x}_{-i},\bs{m},\bs{c})\\
&\stackrel{\text{\ref{assump:7}}}{=} p(Y=y|X_i=x'',\bs{x}_{-i},\bs{m},\bs{c})\\
&= p(y|x'',\bs{x}_{-i},\bs{m},\bs{c}).
\end{align*}
Plugging the derivations for $A$ and $B$ back into Expression \eqref{eq:temp1} we have
\begin{align*}
&= \int_{y,\bs{x}_{-i},\bs{m},\bs{c}} g(y) p(y|x'',\bs{x}_{-i},\bs{m},\bs{c}) p(\bs{m}|x',\bs{x}_{-i},\bs{c})p(\bs{x}_{-i},\bs{c})\\
&= \int_{\bs{x}_{-i},\bs{m},\bs{c}} \E[g(Y)|x'',\bs{x}_{-i},\bs{m},\bs{c}] p(\bs{m}|x',\bs{x}_{-i},\bs{c})p(\bs{x}_{-i},\bs{c}).
\end{align*}
We have
\begin{equation*}
\E[g(Y^{X_i=x'})] \stackrel{\text{\ref{assump:8}}}{=} \E[g(Y^{X_i=x',\bs{M}^{X_i=x'}})] = \int_{\bs{x}_{-i},\bs{m},\bs{c}} \E[g(Y)|x',\bs{x}_{-i},\bs{m},\bs{c}] \underbrace{p(\bs{m}|x',\bs{x}_{-i},\bs{c})p(\bs{x}_{-i},\bs{c})}_{\equiv D}.
\end{equation*}

\bibliography{refs}

\begin{thebibliography}{34}
\providecommand{\natexlab}[1]{#1}
\providecommand{\url}[1]{\texttt{#1}}
\expandafter\ifx\csname urlstyle\endcsname\relax
  \providecommand{\doi}[1]{doi: #1}\else
  \providecommand{\doi}{doi: \begingroup \urlstyle{rm}\Url}\fi

\bibitem[Akbani et~al.(2014)Akbani, Ng, Werner, Shahmoradgoli, Zhang, Ju, Liu,
  Yang, Yoshihara, Li, et~al.]{akbani2014pan}
R.~Akbani, P.~K.~S. Ng, H.~M. Werner, M.~Shahmoradgoli, F.~Zhang, Z.~Ju,
  W.~Liu, J.-Y. Yang, K.~Yoshihara, J.~Li, et~al.
\newblock A pan-cancer proteomic perspective on the cancer genome atlas.
\newblock \emph{Nature communications}, 5\penalty0 (1):\penalty0 1--15, 2014.

\bibitem[Avin et~al.(2005)Avin, Shpitser, and Pearl]{avin2005identifiability}
C.~Avin, I.~Shpitser, and J.~Pearl.
\newblock Identifiability of path-specific effects.
\newblock 2005.

\bibitem[Barfield et~al.(2017)Barfield, Shen, Just, Vokonas, Schwartz,
  Baccarelli, VanderWeele, and Lin]{barfield2017testing}
R.~Barfield, J.~Shen, A.~C. Just, P.~S. Vokonas, J.~Schwartz, A.~A. Baccarelli,
  T.~J. VanderWeele, and X.~Lin.
\newblock Testing for the indirect effect under the null for genome-wide
  mediation analyses.
\newblock \emph{Genetic epidemiology}, 41\penalty0 (8):\penalty0 824--833,
  2017.

\bibitem[Baron and Kenny(1986)]{baron1986moderator}
R.~M. Baron and D.~A. Kenny.
\newblock The moderator--mediator variable distinction in social psychological
  research: Conceptual, strategic, and statistical considerations.
\newblock \emph{Journal of personality and social psychology}, 51\penalty0
  (6):\penalty0 1173, 1986.

\bibitem[Bhattacharyya et~al.(2020)Bhattacharyya, Ha, Liu, Akbani, Liang, and
  Baladandayuthapani]{bhattacharyya2020personalized}
R.~Bhattacharyya, M.~J. Ha, Q.~Liu, R.~Akbani, H.~Liang, and
  V.~Baladandayuthapani.
\newblock Personalized network modeling of the pan-cancer patient and cell line
  interactome.
\newblock \emph{JCO Clinical Cancer Informatics}, 4:\penalty0 399--411, 2020.

\bibitem[Chen and Tsiatis(2001)]{chen2001causal}
P.-Y. Chen and A.~A. Tsiatis.
\newblock Causal inference on the difference of the restricted mean lifetime
  between two groups.
\newblock \emph{Biometrics}, 57\penalty0 (4):\penalty0 1030--1038, 2001.

\bibitem[Davis et~al.(2000)Davis, Solbiati, and
  Cronan]{davis2000overproduction}
M.~S. Davis, J.~Solbiati, and J.~E. Cronan.
\newblock Overproduction of acetyl-coa carboxylase activity increases the rate
  of fatty acid biosynthesis in escherichia coli.
\newblock \emph{Journal of Biological Chemistry}, 275\penalty0 (37):\penalty0
  28593--28598, 2000.

\bibitem[Efron and Tibshirani(1994)]{efron1994introduction}
B.~Efron and R.~J. Tibshirani.
\newblock \emph{An introduction to the bootstrap}.
\newblock CRC press, 1994.

\bibitem[Fasanelli et~al.(2019)Fasanelli, Giraudo, Ricceri, Valeri, and
  Zugna]{fasanelli2019marginal}
F.~Fasanelli, M.~T. Giraudo, F.~Ricceri, L.~Valeri, and D.~Zugna.
\newblock Marginal time-dependent causal effects in mediation analysis with
  survival data.
\newblock \emph{American journal of epidemiology}, 188\penalty0 (5):\penalty0
  967--974, 2019.

\bibitem[Gaynor et~al.(2018)Gaynor, Schwartz, and Lin]{gaynor2018mediation}
S.~M. Gaynor, J.~Schwartz, and X.~Lin.
\newblock Mediation analysis for common binary outcomes.
\newblock \emph{Statistics in medicine}, 2018.

\bibitem[Ha et~al.(2018)Ha, Banerjee, Akbani, Liang, Mills, Do, and
  Baladandayuthapani]{ha2018personalized}
M.~J. Ha, S.~Banerjee, R.~Akbani, H.~Liang, G.~B. Mills, K.-A. Do, and
  V.~Baladandayuthapani.
\newblock Personalized integrated network modeling of the cancer proteome
  atlas.
\newblock \emph{Scientific reports}, 8\penalty0 (1):\penalty0 1--14, 2018.

\bibitem[Hardie(1989)]{hardie1989regulation}
D.~G. Hardie.
\newblock Regulation of fatty acid synthesis via phosphorylation of acetyl-coa
  carboxylase.
\newblock \emph{Progress in lipid research}, 28\penalty0 (2):\penalty0
  117--146, 1989.

\bibitem[Huang and Pan(2016)]{huang2016hypothesis}
Y.-T. Huang and W.-C. Pan.
\newblock Hypothesis test of mediation effect in causal mediation model with
  high-dimensional continuous mediators.
\newblock \emph{Biometrics}, 72\penalty0 (2):\penalty0 402--413, 2016.

\bibitem[Huang et~al.(2014)Huang, VanderWeele, and Lin]{huang2014joint}
Y.-T. Huang, T.~J. VanderWeele, and X.~Lin.
\newblock Joint analysis of snp and gene expression data in genetic association
  studies of complex diseases.
\newblock \emph{The annals of applied statistics}, 8\penalty0 (1):\penalty0
  352, 2014.

\bibitem[Imai et~al.(2010{\natexlab{a}})Imai, Keele, and
  Tingley]{imai2010general}
K.~Imai, L.~Keele, and D.~Tingley.
\newblock A general approach to causal mediation analysis.
\newblock \emph{Psychological methods}, 15\penalty0 (4):\penalty0 309,
  2010{\natexlab{a}}.

\bibitem[Imai et~al.(2010{\natexlab{b}})Imai, Keele, and
  Yamamoto]{imai2010identification}
K.~Imai, L.~Keele, and T.~Yamamoto.
\newblock Identification, inference and sensitivity analysis for causal
  mediation effects.
\newblock \emph{Statistical science}, pages 51--71, 2010{\natexlab{b}}.

\bibitem[Kim et~al.(1989)Kim, Lopez-Casillas, Bai, Luo, and Pape]{kim1989role}
K.-H. Kim, F.~Lopez-Casillas, D.~Bai, X.~Luo, and M.~Pape.
\newblock Role of reversible phosphorylation of acetyl-coa carboxylase in
  long-chain fatty acid synthesis.
\newblock \emph{The FASEB Journal}, 3\penalty0 (11):\penalty0 2250--2256, 1989.

\bibitem[Kumar et~al.(2016)Kumar, Bansal, Narang, Basak, Abbas, and
  Dash]{kumar2016integrating}
D.~Kumar, G.~Bansal, A.~Narang, T.~Basak, T.~Abbas, and D.~Dash.
\newblock Integrating transcriptome and proteome profiling: strategies and
  applications.
\newblock \emph{Proteomics}, 16\penalty0 (19):\penalty0 2533--2544, 2016.

\bibitem[Network et~al.(2013)]{cancer2013comprehensive}
C.~G. A.~R. Network et~al.
\newblock Comprehensive molecular characterization of clear cell renal cell
  carcinoma.
\newblock \emph{Nature}, 499\penalty0 (7456):\penalty0 43, 2013.

\bibitem[Pearl(2001)]{pearl2001direct}
J.~Pearl.
\newblock Direct and indirect effects.
\newblock In \emph{Proceedings of the seventeenth conference on uncertainty in
  artificial intelligence}, pages 411--420. Morgan Kaufmann Publishers Inc.,
  2001.

\bibitem[Pearl(2009)]{pearl2009causality}
J.~Pearl.
\newblock \emph{Causality}.
\newblock Cambridge university press, 2009.

\bibitem[Pearl et~al.(2009)]{pearl2009causal}
J.~Pearl et~al.
\newblock Causal inference in statistics: An overview.
\newblock \emph{Statistics surveys}, 3:\penalty0 96--146, 2009.

\bibitem[Rathmell et~al.(2018)Rathmell, Rathmell, and
  Linehan]{rathmell2018metabolic}
W.~K. Rathmell, J.~C. Rathmell, and W.~M. Linehan.
\newblock Metabolic pathways in kidney cancer: current therapies and future
  directions.
\newblock \emph{Journal of Clinical Oncology}, 36\penalty0 (36):\penalty0
  3540--3546, 2018.

\bibitem[Robins and Greenland(1992)]{robins1992identifiability}
J.~M. Robins and S.~Greenland.
\newblock Identifiability and exchangeability for direct and indirect effects.
\newblock \emph{Epidemiology}, pages 143--155, 1992.

\bibitem[Rubin(1974)]{rubin1974estimating}
D.~B. Rubin.
\newblock Estimating causal effects of treatments in randomized and
  nonrandomized studies.
\newblock \emph{Journal of educational Psychology}, 66\penalty0 (5):\penalty0
  688, 1974.

\bibitem[Sam et~al.(2016)Sam, Teel, Tegge, Bharadwaj, and
  Murali]{sam2016xtalkdb}
S.~A. Sam, J.~Teel, A.~N. Tegge, A.~Bharadwaj, and T.~Murali.
\newblock Xtalkdb: a database of signaling pathway crosstalk.
\newblock \emph{Nucleic acids research}, 45\penalty0 (D1):\penalty0 D432--D439,
  2016.

\bibitem[Tchetgen and Shpitser(2012)]{tchetgen2012semiparametric}
E.~J.~T. Tchetgen and I.~Shpitser.
\newblock Semiparametric theory for causal mediation analysis: efficiency
  bounds, multiple robustness, and sensitivity analysis.
\newblock \emph{Annals of Statistics}, 40\penalty0 (3):\penalty0 1816, 2012.

\bibitem[Uno et~al.(2014)Uno, Claggett, Tian, Inoue, Gallo, Miyata, Schrag,
  Takeuchi, Uyama, Zhao, et~al.]{uno2014moving}
H.~Uno, B.~Claggett, L.~Tian, E.~Inoue, P.~Gallo, T.~Miyata, D.~Schrag,
  M.~Takeuchi, Y.~Uyama, L.~Zhao, et~al.
\newblock Moving beyond the hazard ratio in quantifying the between-group
  difference in survival analysis.
\newblock \emph{Journal of clinical Oncology}, 32\penalty0 (22):\penalty0 2380,
  2014.

\bibitem[Valeri et~al.(2014)Valeri, Lin, and VanderWeele]{valeri2014mediation}
L.~Valeri, X.~Lin, and T.~J. VanderWeele.
\newblock Mediation analysis when a continuous mediator is measured with error
  and the outcome follows a generalized linear model.
\newblock \emph{Statistics in medicine}, 33\penalty0 (28):\penalty0 4875--4890,
  2014.

\bibitem[VanderWeele and Vansteelandt(2009)]{vanderweele2009conceptual}
T.~J. VanderWeele and S.~Vansteelandt.
\newblock Conceptual issues concerning mediation, interventions and
  composition.
\newblock \emph{Statistics and its Interface}, 2\penalty0 (4):\penalty0
  457--468, 2009.

\bibitem[VanderWeele and Vansteelandt(2010)]{vanderweele2010odds}
T.~J. VanderWeele and S.~Vansteelandt.
\newblock Odds ratios for mediation analysis for a dichotomous outcome.
\newblock \emph{American journal of epidemiology}, 172\penalty0 (12):\penalty0
  1339--1348, 2010.

\bibitem[VanderWeele et~al.(2014)VanderWeele, Vansteelandt, and
  Robins]{vanderweele2014effect}
T.~J. VanderWeele, S.~Vansteelandt, and J.~M. Robins.
\newblock Effect decomposition in the presence of an exposure-induced
  mediator-outcome confounder.
\newblock \emph{Epidemiology (Cambridge, Mass.)}, 25\penalty0 (2):\penalty0
  300, 2014.

\bibitem[Wei et~al.(2017)Wei, Jin, Yang, Xu, Zhu, and
  Ji]{wei2017tcgawei2017tcga}
L.~Wei, Z.~Jin, S.~Yang, Y.~Xu, Y.~Zhu, and Y.~Ji.
\newblock Tcga-assembler 2: software pipeline for retrieval and processing of
  tcga/cptac data.
\newblock \emph{Bioinformatics}, 34\penalty0 (9):\penalty0 1615--1617, 2017.

\bibitem[Zhao et~al.(2020)Zhao, Lindquist, and Caffo]{zhao2020sparse}
Y.~Zhao, M.~A. Lindquist, and B.~S. Caffo.
\newblock Sparse principal component based high-dimensional mediation analysis.
\newblock \emph{Computational Statistics \& Data Analysis}, 142:\penalty0
  106835, 2020.

\end{thebibliography}

\end{document}

\begin{landscape}
\noindent
\textbf{Supplementary Material for ``A Framework for Mediation Analysis with Multiple Exposures, Multivariate Mediators, and Non-Linear Response Models''}
% latex table generated in R 3.5.1 by xtable 1.8-3 package
% Tue Aug 25 10:00:36 2020
\begin{table}[ht]
\centering
\begin{tabular}{rrlrlrlrlrl}
  \hline
  Pathway & \multicolumn{2}{c}{PTEN}& \multicolumn{2}{c}{TCA cycle}& \multicolumn{2}{c}{FAS}& \multicolumn{2}{c}{AMPK}& \multicolumn{2}{c}{Pentose} \\
 \hline
Apoptosis & 24 & [-39,89] & 42 & [-25,120] & 34 & [-75,144] & -32 & [-93,53] & 33 & [-104,221] \\ 
  Cell cycle & -7 & [-61,78] & 76 & [-16,164] & 11 & [-111,129] & -36 & [-136,47] & 18 & [-162,212] \\ 
  DNA damage response & -6 & [-70,95] & 34 & [-25,127] & -125 & [-292,-21] & -17 & [-125,98] & -171 & [-313,22] \\ 
  EMT & 0 & [-45,47] & 52 & [-13,140] & -87 & [-186,15] & -11 & [-93,69] & -140 & [-271,-10] \\ 
  Hormone receptor & -4 & [-90,88] & 47 & [-25,153] & -66 & [-213,47] & 48 & [-81,174] & 108 & [-62,284] \\ 
  Hormone signaling (Breast) & 4 & [-34,68] & 34 & [-18,77] & 23 & [-76,110] & -40 & [-106,23] & 9 & [-119,139] \\ 
  PI3K/AKT & 48 & [-80,141] & 135 & [38,258] & -26 & [-135,71] & 20 & [-88,148] & 20 & [-148,237] \\ 
  RAS/MAPK & 6 & [-59,86] & 106 & [12,231] & -65 & [-188,78] & 3 & [-103,140] & 5 & [-186,206] \\ 
  RTK & 29 & [-35,109] & 48 & [-25,115] & -99 & [-229,5] & 126 & [-31,312] & -78 & [-238,122] \\ 
  TSC/mTOR & 27 & [-56,150] & 157 & [33,307] & -131 & [-291,5] & -46 & [-172,87] & -52 & [-217,164] \\ 
  Breast reactive & -34 & [-75,28] & 98 & [31,197] & -30 & [-114,51] & -8 & [-81,67] & -5 & [-149,135] \\ 
  Core reactive & -33 & [-77,21] & 111 & [23,205] & 4 & [-104,70] & 25 & [-51,133] & 31 & [-108,195] \\ 
   \hline
\end{tabular}
\caption{Indirect Effects in days of metabolomic mRNA expression as mediated by protein pathways.} 
\end{table}

\end{landscape}